\documentclass[aps, prd, twocolumn, nofootinbib]{revtex4}

\usepackage{amsfonts,amsmath,amsthm,amssymb,graphicx}

\begin{document}

\title{Non-canonical generalizations of slow-roll inflation models}

\author{Konstantinos Tzirakis} 
\email{ct38@buffalo.edu}
\author{William H. Kinney} 
\email{whkinney@buffalo.edu}
\affiliation{Dept. of Physics, University at Buffalo, the  State University of New York, Buffalo, NY 14260-1500}

\begin{abstract}
We consider non-canonical generalizations of two classes of single-field inflation models. First, we study the non-canonical version of ``ultra-slow roll'' inflation, which is a class of inflation models for which quantum modes do not freeze at horizon crossing, but instead evolve rapidly on superhorizon scales. Second, we consider the non-canonical generalization of the simplest ``chaotic'' inflation scenario, with a potential dominated by a quadratic (mass) term for the inflaton.  We find a class of related non-canonical solutions with polynomial potentials, but with varying speed of sound. These solutions are characterized by a constant field velocity, and we dub such models {\it isokinetic} inflation. As in the canonical limit, isokinetic inflation has a slightly red-tilted power spectrum, consistent with current data. Unlike the canonical case, however, these models can have an arbitrarily small tensor/scalar ratio. Of particular interest is that isokinetic inflation is marked by a correlation between the tensor/scalar ratio and the amplitude  of non-Gaussianity such that parameter regimes with small tensor/scalar ratio have {\it large} associated non-Gaussianity, which is a distinct observational signature. 
\end{abstract}

\maketitle

\section{Introduction}
\label{Introduction}

Inflation \cite{Guth:1980zm,Linde:1981mu,Albrecht:1982wi} has emerged as the most promising framework for understanding the physics of the very early universe. In particular, it ties the evolution of the universe to the properties of one or more scalar {\it inflaton} fields, which are responsible for creating a period of rapidly accelerating expansion. This period of acceleration naturally creates a flat, homogeneous patch which later evolves into the present universe. Inflation makes detailed predictions, most importantly the generation of primordial density and gravitational-wave perturbations \cite{Mukhanov:1981xt,Hawking:1982my,Starobinsky:1982ee,Guth:1982ec,Bardeen:1983qw,Starobinsky:1979ty,Starobinsky:1980te} which are an excellent match to current high-precision data. However, the identity of the inflaton field is currently unknown, except that it must lie outside the Standard Model for particle physics. A promising place to search for a fundamental theory of inflation is within the ``landscape'' of string theory, which predicts a plethora of scalar fields associated with the compactification of extra dimensions and the configuration of lower-dimensional ``branes'' moving in a higher-dimensional bulk space. Recent developments in string theory have produced a number of phenomenologically viable stringy inflation models such as the KKLMMT scenario \cite{Kachru:2003sx}, Racetrack Inflation \cite{BlancoPillado:2004ns}, and Roulette Inflation \cite{Bond:2006nc}. The Dirac-Born-Infeld (DBI) scenario \cite{Silverstein:2003hf} has attracted particular interest because of the novel feature that slow roll can be achieved through a low sound speed instead of from dynamical friction due to expansion.  An important lesson from these recent theoretical developments is that it is not necessarily safe to assume that the inflaton is a single, canonical scalar field. Instead, a varying sound speed arises naturally within many string-inspired inflation scenarios. 

The corresponding observational picture is similarly broadened by these developments in model building: canonical slow-roll inflation models generically predict a strong suppression of scale-dependence and non-Gaussianity in the perturbation spectra generated during inflation. Therefore, the observables most important for constraining such scenarios are the scalar spectral index $n_{s}$ and the tensor/scalar ratio $r$ \cite{Dodelson:1997hr,Kinney:1998md}. Current data constrain the spectral index well enough to rule out a number of well-motivated inflationary scenarios, but there is as of yet no evidence for non-vanishing $r$ \cite{Spergel:2006hy,Alabidi:2006qa,Seljak:2006bg,Kinney:2006qm,Martin:2006rs,Komatsu:2008hk,Dunkley:2008ie,Kinney:2008wy,Lorenz:2008je,Lorenz:2008et}. Inflation from non-canonical Lagrangians introduces novel phenomenology above and beyond this simple set of observables, in particular significant non-Gaussianity \cite{Alishahiha:2004eh,Chen:2006nt,Spalinski:2007qy,Bean:2007eh,LoVerde:2007ri}. This creates a potentially powerful new ``window'' for probing inflationary physics. This is particularly significant for phenomenological efforts to ``reconstruct'' the inflationary potential \cite{Lidsey:1995np}, since non-canonical models present two free functions, the scalar field potential and the speed of sound, which must be reconstructed. The canonical ``flow'' formalism for inflation \cite{Liddle:1994dx,Kinney:2002qn,Schwarz:2001vv} has been generalized to the case of DBI inflation \cite{Peiris:2007gz,Lorenz:2008je} as well as to the case of a completely general Lagrangian \cite{Bean:2008ga,Agarwal:2008ah}.

In this paper, we consider the non-canonical generalizations of two interesting classes of canonical inflation models. First, we study the non-canonical generalization of ``ultra-slow roll'' inflation, which is a class of inflation models for which the slow-roll approximation is strongly violated, so that quantum modes evolve rapidly on superhorizon scales \cite{Kinney:2005vj,Tzirakis:2007bf}. Background solutions for this case were calculated by Spalinski \cite{Spalinski:2007un}. Here, we fully calculate the associated fluctuation power spectra. Second, we consider the non-canonical generalization of the simplest ``chaotic'' inflation scenario, with a potential $V\left(\phi\right) = m^2 \phi^2$, for which the field evolves with approximately constant velocity $\dot\phi \simeq {\mathrm const.}$  We find a class of related non-canonical solutions with polynomial potentials $V\left(\phi\right) \propto \phi^p$ and constant field velocity, but with varying speed of sound, which we dub {\it isokinetic} inflation. Unlike the canonical case, the non-canonical model can have an arbitrarily small tensor/scalar ratio combined with a slightly red-tilted power spectrum, $1 - n \sim 0.05$, consistent with current data. Of particular interest is that this class of models is marked by a correlation between the tensor/scalar ratio and the amplitude $f_{NL}$ of non-Gaussianity such that parameter regimes with small tensor/scalar ratio have {\it large} associated non-Gaussianity. 

In the next section, we review the calculation of cosmological observables in canonical and non-canonical inflation models. 

\section{Cosmological Observables}
\label{sec: Cosmological Observables}

\subsection{Inflation from canonical scalar fields}
\label{sec: Inflation from canonical scalar fields}

Inflation can be easily implemented in scalar field models. Consider the following action of a real scalar field $\phi$ 
\begin{equation}
\label{eq: action for canonical}
S=\int d^4x\sqrt{-g}\left[\frac{1}{2}g^{\mu\nu}{\partial_\mu \phi}{\partial_\nu \phi}-V(\phi)\right],
\end{equation}
with a flat Friedmann-Robertson-Walker (FRW) metric 
\begin{equation}
\label{eq: background metric}
ds^2 = dt^2-a^2(t)d \mathbf{x}^2=a^2(\tau)[d\tau^2-d \mathbf{x}^2].
\end{equation}
It can then be shown that the equation of motion for the scalar field is given by the equation
\begin{equation}
\label{eq: equation of motion for canonical}
\ddot \phi + 3 H \dot \phi + V'(\phi)=0,
\end{equation}
where the conformal time $\tau$ can be expressed in terms of the coordinate time $t$ as 
\begin{equation}
\label{eq: tau}
d\tau=\frac{1}{a}dt,
\end{equation}
and the Hubble parameter $H$ is defined to be
\begin{equation}
\label{eq: H}
H \equiv \frac{\dot a(t)}{a(t)}.
\end{equation}
Unless otherwise stated, overdots denote derivatives with respect to the coordinate time $t$, and primes derivatives with respect to the field $\phi$. If the coordinate time $t$ is a single-valued function of the inflaton field $\phi$, we can equivalently express the Hubble parameter $H$ in  terms of $\phi$ instead of $t$. The  dynamics of single  field  inflation  can  then  be  described  by  the  so-called {\it Hamilton-Jacobi}  formalism   \cite{Muslimov:1990be,   Salopek:1990jq, Lidsey:1995np}
\begin{eqnarray}
\label{eq: HJ for canonical} 
\left[H^{\prime}  (\phi)\right]^2-\frac{3}{2M_{P}^2}H^2(\phi)  &=& -\frac{1}{2M_{P}^4}V(\phi),\cr
\dot{\phi}=-2M_{P}^2H'(\phi).&&
\end{eqnarray}
This system of  two first order differential equations is  equivalent to the second order equation of motion (\ref{eq: equation of motion for canonical}). Since we now use the field $\phi$ as our clock, we can define the following Hubble slow roll parameters in terms of derivatives of $H$ with respect to $\phi$ as \cite{Kinney:2002qn}
\begin{eqnarray}
\label{eq: intro flow 1}
\epsilon &\equiv& 2M_{P}^2\left(\frac{H'(\phi)}{H(\phi)}\right)^2, \cr
\eta &\equiv& 2M_{P}^2\frac{H''(\phi)}{H(\phi)}, \cr &\vdots& \cr
{}^\ell \lambda\left(\phi\right) &\equiv& \left(2 M_P^2\right)^{\ell} \left(\frac{H'\left(\phi\right)}{H\left(\phi\right)}\right)^{\ell - 1} \frac{1}{H\left(\phi\right)} \frac{d^{\ell + 1} H\left(\phi\right)}{d \phi^{\ell + 1}},
\end{eqnarray}
where $\ell=2,...,\infty$ is an integer index. Since the above tower of flow parameters is defined in terms of derivatives of $H$, we will refer to it as the {\it $H$-tower}. The potential $V(\phi)$ is then given by the following exact expression
\begin{equation}
\label{eq: V for canonical}
V(\phi)=3M_{P}^2H^2(\phi)\left(1-\frac{1}{3}\epsilon(\phi)\right).
\end{equation}
As long as the above parameters are varying slowly, the power spectra of curvature perturbations $P_{\mathcal{R}}$ and of tensor perturbations $P_{T}$ are given by
\begin{eqnarray}
\label{eq: spectra 1}
P_{\mathcal{R}}&=&\frac{1}{8\pi^2}\left.\frac{H^2}{M_{P}^2\epsilon}\right\vert_{k=aH}, \cr P_{T}&=&\frac{2}{\pi^2}\left.\frac{H^2}{M_{P}^2}\right\vert_{k=aH}.
\end{eqnarray}

We then define the tensor/scalar ratio to lowest order in slow roll as
\begin{equation}
\label{eq: r for canonical}
r \equiv \frac{P_{T}}{P_{\mathcal{R}}}=16 \epsilon.
\end{equation}
If we also approximate the power spectra by power laws \cite{Kinney:2005in}
\begin{eqnarray}
\label{eq: spectra 2}
P_{\mathcal{R}}(k) &\propto& k^{n_{s}-1}, \cr P_{T}(k) &\propto& k^{n_{T}},
\end{eqnarray}
we can express the spectral indices to lowest order in slow roll by the following equations
\begin{eqnarray}
\label{eq: spectral indices 1}
n_{s}-1\equiv \frac{d{\rm ln}P_{\mathcal{R}}}{d{\rm ln}k} &=& -4\epsilon+2\eta, \cr n_{T} \equiv \frac{d{\rm ln}P_{T}}{d{\rm ln}k}&=& -2\epsilon.
\end{eqnarray}
There are therefore three independent parameters that can describe any inflationary model to lowest order, $P_{\mathcal{R}}$, $P_{T}$ and $n_{s}$. Another independent observable is the signature of non-Gaussian features of the primordial perturbations, but it has conclusively been shown \cite{Maldacena:2002vr,Acquaviva:2002ud} that it is negligible in slow-roll inflation models.

\subsection{D-brane Inflation}
\label{sec: Inflation from warped D-branes}

In this section we review the framework of D-brane inflation. In most cases it evolves a mobile D3-brane which moves in a six-dimensional ``throat'' with the following metric: \cite{Klebanov:2000hb}
\begin{equation}
\label{eq: DBImetric}
ds^2_{10} = h^2\left(r\right) ds^2_4 + h^{-2}\left(r\right) \left(d r^2 + r^2 ds^2_{X_5}\right),
\end{equation}
where the manifold $X_{5}$ shapes the base of the cone. The effective potential is then the result of the interaction between the D3-brane and the anti-D3-branes which can either be located at the tip of the throat ({\it UV} model) \cite{Chen:2006nt,Spalinski:2007qy,Bean:2007eh,LoVerde:2007ri,Lidsey:2007gq}, or at the infrared ends of different throats ({\it IR} model) \cite{Chen:2004gc,Chen:2005ad}.

One of the scenarios that appears to be promising in the quest for an explicit realization of the inflationary paradigm from superstring theory is the DBI scenario, because of the new phenomenology that it introduces. In the DBI scenario, the inflaton field $\phi$ is the degree of freedom that is associated with the D3-brane and is simply given by the rescaled coordinate in the throat $r$ between the brane and the antibranes
\begin{equation}
\label{eq: field for DBI}
\phi=\sqrt{T_{3}}r,
\end{equation}
where $T_{3}$ is the brane tension. The dynamics of the D3-brane are then determined by the DBI action
\begin{eqnarray}
\label{eq: DBI action}
&&S=-\int d^4x\sqrt{-g}\cr &&\left[ f^{-1}\left(\phi\right) \left(\sqrt{1 + f\left(\phi\right) g^{\mu\nu} \partial_\mu \phi \partial_\nu \phi} - 1\right) + V\right],
\end{eqnarray}
where $V=V\left(\phi\right)$ is an arbitrary potential, and $f^{-1}\left(\phi\right)$ is the brane tension expressed in terms of the warp factor $h\left(\phi\right)$ as
\begin{equation}
\label{eq: warpfactor}
f^{-1}\left(\phi\right) = T_{3}h^4\left(\phi\right).
\end{equation}
If we again assume a four-dimensional metric of the FRW form, the equation of motion for the inflaton field $\phi$ can then be written as
\begin{equation}
\label{eq: EOM for DBI}
\ddot \phi +\frac{3H \dot \phi}{\gamma^2}+\frac{V'}{\gamma^3}+\frac{3f'}{2f}\dot \phi^2+\frac{f'}{f^2}\left(\frac{1}{\gamma^3}-1\right)=0.
\end{equation}
The Lorentz factor $\gamma=\gamma(\phi)$, is given by 
\begin{equation}
\label{eq: gamma for DBI}
\gamma=\frac{1}{\sqrt{1-f(\phi)\dot \phi^2}},
\end{equation}
and can result in many e-folds of inflation, even if the potential is steep. We next define the speed of sound $c_{s}$ as the speed at which the fluctuations of the inflaton field propagate relative to the homogeneous background as
\begin{equation}
\label{eq: cs for DBI 2}
c_{s}^2=\frac{dP}{d\rho},
\end{equation}
where $P$ and $\rho$ are the pressure and energy density of the field. We can then express $\gamma$ in terms of $c_{s}$ as follows:
\begin{equation}
\label{eq: gamma for DBI 2}
\gamma^2=\frac{1}{c_{s}^2}.
\end{equation}
In the DBI scenario therefore, slow roll is not achieved from dynamical friction due to the expansion, but rather through a low speed of sound.

Following the analysis of section \ref{sec: Inflation from canonical scalar fields}, we can define the Hamilton-Jacobi formalism for DBI as \cite{Spalinski:2007kt}
\begin{eqnarray}
\label{eq:Hamjacobi for DBI 1}
3M_P^2H^2(\phi)-V(\phi)&=&\frac{\gamma(\phi)-1}{f(\phi)},\cr
\dot \phi=-\frac{2M_P^2}{\gamma(\phi)}H'(\phi),&&
\end{eqnarray}
where
\begin{equation}
\label{eq: gamma H'}
\gamma=\sqrt{1+4M_{P}^2f(\phi)H'(\phi)}.
\end{equation}
In addition to the $H$-tower defined previously, we can also define an infinite {\it $\gamma$-tower} of flow parameters that are expressed in terms of derivatives of $\gamma$ as follows \cite{Peiris:2007gz}:
\begin{eqnarray}
\label{eq:flowparams}
\epsilon\left(\phi\right) &\equiv& \frac{2 M_P^2}{\gamma\left(\phi\right)} \left(\frac{H'\left(\phi\right)}{H\left(\phi\right)}\right)^2,\cr
\eta\left(\phi\right) &\equiv& \frac{2 M_P^2}{\gamma\left(\phi\right)} \frac{H''\left(\phi\right)}{H\left(\phi\right)},\cr &\vdots& \cr
{}^\ell \lambda\left(\phi\right) &\equiv& \left(\frac{2 M_P^2}{\gamma\left(\phi\right)}\right)^{\ell} \left(\frac{H'\left(\phi\right)}{H\left(\phi\right)}\right)^{\ell - 1} \frac{1}{H\left(\phi\right)} \frac{d^{\ell + 1} H\left(\phi\right)}{d \phi^{\ell + 1}},\cr
s\left(\phi\right) &\equiv& \frac{2 M_P^2}{\gamma\left(\phi\right)}\frac{H'\left(\phi\right)}{H\left(\phi\right)}\frac{\gamma'\left(\phi\right)}{\gamma\left(\phi\right)},\cr
\rho\left(\phi\right) &\equiv& \frac{2 M_P^2}{\gamma\left(\phi\right)}  \frac{\gamma''\left(\phi\right)}{\gamma\left(\phi\right)},\cr &\vdots& \cr
{}^\ell \alpha\left(\phi\right) &\equiv& \left(\frac{2 M_P^2}{\gamma\left(\phi\right)}\right)^{\ell} \left(\frac{H'\left(\phi\right)}{H\left(\phi\right)}\right)^{\ell - 1} \frac{1}{\gamma\left(\phi\right)} \frac{d^{\ell + 1} \gamma\left(\phi\right)}{d \phi^{\ell + 1}}.
\end{eqnarray}
The potential can then be written as
\begin{equation}
\label{eq: V for BDI}
V(\phi)=3M_{P}^2H^2(\phi)\left(1-\frac{2\epsilon(\phi)}{3}\frac{\gamma(\phi)}{\gamma(\phi)+1}\right).
\end{equation}
To lowest order, the power spectra of curvature perturbations $P_{\mathcal{R}}$ and of tensor perturbations $P_{T}$ are given respectively by
\begin{eqnarray}
\label{eq: spectra for DBI 1}
P_{\mathcal{R}}&=&\frac{1}{8 \pi^2}\left.\frac{H^2}{c_{s}\epsilon M_{P}^2}\right\vert_{c_{s}k=aH}, \cr P_{T}&=&\frac{2}{\pi^2}\left.\frac{H^2}{M_{P}^2}\right\vert_{k=aH},
\end{eqnarray}
and then
\begin{equation}
\label{eq: r for DBI}
r \equiv \frac{P_{T}}{P_{\mathcal{R}}}=16 c_{s}\epsilon.
\end{equation}
The spectral indices can then be described by the following expressions to lowest order in the flow parameters
\begin{eqnarray}
\label{eq: spectral indices for DBI 1}
n_{s}-1\equiv \frac{d{\rm ln}P_{\mathcal{R}}}{d{\rm ln}k} &=& -4\epsilon+2\eta-2s, \cr n_{T} \equiv \frac{d{\rm ln}P_{T}}{d{\rm ln}k}&=& -2\epsilon,
\end{eqnarray}
where the scalar spectral index now depends on the variation of the speed of sound as expressed by the flow parameter $s$.

What makes the DBI scenario very attractive is the novel characteristic of significant non-Gaussianities \cite{Chen:2006nt,Spalinski:2007qy,Bean:2007eh,LoVerde:2007ri,Alishahiha:2004eh}. It has been shown \cite{Chen:2006nt} that the sound speed is related to the primordial non-Gaussianity by
\begin{equation}
\label{eq: fnl}
f_{NL}^{\rm equil.} = \frac{35}{108} \left(\frac{1}{c_s^2} -1 \right).
\end{equation}
A direct measurement of $f_{NL}$ can therefore be used not only to distinguish the DBI scenario from other candidates, but also to constrain the speed of sound while the CMB scales were exiting the sound horizon. 

In the limit obviously where
\begin{equation}
\label{eq: bridge}
c_{s}\rightarrow 1 \hspace{0.3cm} {\rm and} \hspace{0.3cm} s\rightarrow 0,
\end{equation}
the above analysis reduces to the case of a canonical scalar field presented in section \ref{sec: Inflation from canonical scalar fields}. In the next section we will study the evolution of quantum modes in the general case of non-constant speed of sound. One can always recover the canonical case by simply using Eq. (\ref{eq: bridge}) in all of the following results.

\section{Solution for quantum modes and the horizon crossing formalism}
\label{sec:Solution for quantum modes and the horizon crossing formalism}

One of the most attractive aspects of the inflationary paradigm is that it provides a natural explanation for the large-scale structure of the universe. It is based on the property of the inflationary spacetimes which stretch quantum fluctuations from extremely small scales to superhorizon scales during inflation. These fluctuations are encoded as metric perturbations, which during inflation are of two types: scalar (or curvature) perturbations, and tensor perturbations (or gravitational waves).

A distinctive feature of models which are described by non-canonical Lagrangians is the fact that the fluctuations of the inflaton field do not propagate with the speed of light, but rather with the speed of sound $c_{s}$ which can differ from unity. As a result, scalar perturbations are sensitive to the sound horizon $c_{s}H^{-1}$, whereas tensor perturbations are not. A detailed treatment of the generation of perturbations for an arbitrary speed of sound was done by Garriga and Mukhanov \cite{Garriga:1999vw}, who showed that the power spectrum of curvature perturbations is given by
\begin{equation}
\label{eq: scalar spectrum}
P_{\cal R}\left(k\right) = \frac{k^3}{2 \pi^2} \left\vert\frac{u_k}{z}\right\vert^2,
\end{equation}
where the quantum mode function $u_k$ satisfies
\begin{equation}
\label{eq:uk}
u_k'' + \left(c_s^2 k^2 - \frac{z''}{z}\right) u_k = 0,
\end{equation}
and $z$ is defined by
\begin{equation}
z \equiv \frac{a \sqrt{\rho + p}}{c_s H} = \frac{a \gamma^{3/2} \dot\phi}{H} = - M_P a \gamma \sqrt{2 \epsilon}.
\end{equation}
In the above expressions, a prime denotes a derivative with respect to the conformal time $\tau$ and an overdot denotes a derivative with respect to the coordinate time $t$.

We can equivalently express Eq. (\ref{eq: scalar spectrum}) in terms of the time variable $y$, which is given by
\begin{equation}
\label{eq:y}
y=\frac{c_{s}k}{aH}=\frac{k}{\gamma a H},
\end{equation}
and is the ratio of the wavelength of the mode relative to the sound horizon size \cite{Kinney:2007ag}. The mode equation then becomes
\begin{equation}
\label{eq:exactmode}
\left(1 - \epsilon -s\right)^2 y^2 \frac{d^2 u_k}{dy^2} + G y \frac{d u_k}{d y} + \left[y^2 - F\right] u_k = 0,
\end{equation}
where
\begin{equation}
\label{eq:G}
G = -s + 3 \epsilon s - 2 \epsilon \eta - \eta s + 2\epsilon^2 + 3 s^2 - \epsilon \rho,
\end{equation}
and
\begin{eqnarray}
\label{eq:F}
F &=&2 + 2 \epsilon - 3 \eta - \frac{3}{2} s - 4 \epsilon \eta + \frac{1}{2} \eta s  + 2 \epsilon^2\cr
 &&+ \eta^2 - \frac{3}{4} s^2 + {}^2\lambda + \frac{1}{2}\epsilon\rho,
\end{eqnarray}
are exact functions of the flow parameters. 

In order to evaluate the scalar power spectrum $P_{\mathcal{R}}(k)$, we need to solve the mode equation and evaluate the quantity $\left\vert u_{k}/z \right\vert$ for every mode with comoving wavenumber $k$. Standard lore has it that quantum fluctuations during inflation ``freeze out'' (cease to evolve) after horizon exit and they behave as classical perturbations thereafter \cite{Mukhanov:1981xt,Hawking:1982my,Starobinsky:1982ee,Guth:1982ec,Bardeen:1983qw}. We can therefore calculate $P_{\mathcal{R}}(k)$ at horizon crossing ($y=1$) rather than the long wavelength limit ($y \rightarrow 0$). Even though this is exactly true for power-law inflation and approximately true for slow roll in the case of canonical scalar fields, WHK showed in \cite{Kinney:2005vj} that one must be careful when using the horizon crossing formalism. There are viable models for which the horizon crossing formalism is invalid, and one is forced to evaluate the power spectrum in the long wavelength limit. In the remaining of this section we will generalize the above result of \cite{Kinney:2005vj} to the more general case where the speed of sound is an arbitrary function of time. It should be noted at this point that since the quantum fluctuations become superhorizon when they exit the sound horizon, we will use the more appropriate term ``sound horizon crossing'' formalism, instead of ``horizon crossing'' formalism used for the case of canonical scalar fields.

We first solve the case where all the flow parameters are constant (see \cite{Kinney:2007ag} for details), and we then generalize for an arbitrary flow evolution. If we define $N$ to be the number of e-folds before the end of inflation, then
\begin{equation}
\label{eq:numefolds}
N \equiv - \int{H}{dt} = \frac{1}{\sqrt{2 M_P^2}} \int_{\phi_{end}}^{\phi}{\sqrt{\frac{\gamma\left(\phi\right)}{\epsilon\left(\phi\right)}} d\phi},
\end{equation}
so that $N = 0$ at the end of inflation and increases as one goes backwards in time. Since the flow parameters are taken to be constant, we can find expressions for the derivatives of the flow parameters (\ref{eq:flowparams}) with respect to $N$, and set them to zero
\begin{equation}
\label{eq: deriv of flow params are zero}
\frac{d \epsilon}{d N} = \frac{d {}^\ell \lambda}{d N} =\frac{d s}{d N} = \frac{d {}^\ell \alpha}{d N} = 0.
\end{equation}
We then find that 
\begin{eqnarray}
\label{eq:flow parameters}
\eta&=&\frac{1}{2}\left(2\epsilon+s\right),\cr \xi&=&\frac{1}{2}\left(2\epsilon+s\right)\left(\epsilon+s\right),\cr \rho&=&\frac{3s^2}{2\epsilon},
\end{eqnarray}
and the mode equation reduces to 
\begin{eqnarray}
\label{eq:mode equation for constant parms. 1}
(&1-\epsilon-s&)^2y^2\frac{d^2u_{k}}{dy^2}+s(-1+\epsilon+s)y\frac{du_{k}}{dy}\cr&&+[y^2-(1-s)(2-s-\epsilon)]u_{k}=0.
\end{eqnarray}
The solutions of Eq. (\ref{eq:mode equation for constant parms. 1}) are proportional to Hankel functions and after applying the choice of vacuum and the Wronskian condition for the modes, we obtain the following normalized solution
\begin{equation}
\label{eq:solution of mode eqaution for constant parms. 3}
u_{k}(y)=\frac{1}{2}\sqrt{\frac{\pi}{c_{s}k}}\sqrt{\frac{y}{1-\epsilon-s}}H_{\nu}\left(\frac{y}{1-\epsilon-s}\right),
\end{equation}
where
\begin{equation}
\label{eq:nu for mode equation for constant parms. 1}
\nu= \frac{3-2s-\epsilon}{2(1-\epsilon-s)}.
\end{equation}
In the long wavelength limit the above solution takes the asymptotic form
\begin{equation}
\label{eq:solution of mode eqaution for constant parms. at small y}
\left|u_{k}(y)\right| \rightarrow 2^{\nu-3/2}\frac{\Gamma(\nu)}{\Gamma(3/2)}(1-\epsilon-s)^{\nu-1/2} \frac{y^{1/2-\nu}}{\sqrt{2 c_{s}k}},
\end{equation}
and if we set
\begin{equation}
\label{eq:V(v)}
\mathcal{V}(\nu)=2^{\nu-3}\frac{\Gamma(\nu)}{\Gamma(3/2)}(1-\epsilon-s)^{\nu-1/2},
\end{equation}
we finally find that
\begin{equation}
\label{eq:power spectrum a}
P_{\mathcal{R}}^{1/2}=\left(\frac{\mathcal{V}(\nu)}{\pi M_{P}}\right) \frac{H}{\sqrt{c_{s}\epsilon}}y^{3/2-\nu}.
\end{equation}
The sound horizon crossing formalism can then equivalently be expressed by the condition \cite{Kinney:2005vj} 
\begin{equation}
\label{eq:y formalism 1}
\frac{d}{dy}\left(\frac{H}{\sqrt{c_{s}\epsilon}}y^{3/2-\nu}\right)=0,
\end{equation}
which states that the quantity inside the parenthesis does not depend on $y$ and therefore we can evaluate the power spectrum at any preferred $y$ value, which is conventionally chosen to be the sound horizon crossing for which $y=1$.

It is now obvious that for different flow evolutions, the condition (\ref{eq:y formalism 1}) may only be approximately true or even strongly violated in some cases, and one should seek a generalization of it for the case of an arbitrary evolution. This can be done by first noticing that 
\begin{equation}
\label{eq:y formalism 2}
\frac{d}{dy}=-\frac{1}{c_{s}k(1-\epsilon-s)}\frac{d}{d\tau}=\frac{1}{y(1-\epsilon-s)}\frac{d}{dN},
\end{equation}
exactly. We can then write
\begin{eqnarray}
\label{eq:y formalism 3}
y\frac{d}{dy}\left(\frac{H}{\sqrt{c_{s}\epsilon}}y^{3/2-\nu}\right)&&=\left(\frac{3}{2}-\nu\right)\left(\frac{H}{\sqrt{c_{s}\epsilon}}y^{3/2-\nu}\right)\cr && +\frac{y^{3/2-\nu}}{1-\epsilon-s}\frac{d}{dN}\left(\frac{H}{\sqrt{c_{s}\epsilon}}\right).
\end{eqnarray}
Using finally that 
\begin{equation}
\label{eq:sqrteps dN}
\frac{d}{dN}\left(\frac{1}{\sqrt{\epsilon}}\right)=\frac{1}{2\sqrt{\epsilon}}(2\epsilon+s-2\eta),
\end{equation}
and 
\begin{equation}
\label{eq:sqrtcs dN}
\frac{d}{dN}\left(\frac{1}{\sqrt{c_{s}}}\right)=\frac{s}{2\sqrt{c_{s}}},
\end{equation}
we recover the following expression for an arbitrary evolution
\begin{equation}
\label{eq:y formalism 4}
\frac{d}{d{\rm ln}y}\left[{\rm ln}\left(\frac{H}{\sqrt{c_{s}\epsilon}}y^{3/2-\nu}\right)\right]=\frac{3}{2}-\nu+\frac{2\epsilon-\eta+s}{1-\epsilon-s}.
\end{equation}
It is straightforward to show that the above result vanishes in the de Sitter case where $\epsilon=\eta=s=0$, and in the case where all the flow parameters are constant.

The generalization to slow roll is trivial. During slow roll the order of the Hankel function will not be given by Eq. (\ref{eq:nu for mode equation for constant parms. 1}), but rather from the following expression
\begin{equation}
\label{eq:nu for mode equation for constant parms. 1a}
\nu=\frac{3}{2}+2\epsilon-\eta+s.
\end{equation}
Equation (\ref{eq:y formalism 4}) then takes the form
\begin{eqnarray}
\label{eq:y formalism 5}
\frac{d}{d{\rm ln}y}&&\left[{\rm ln}\left(\frac{H}{\sqrt{c_{s}\epsilon}}y^{3/2-\nu}\right)\right]=-2\epsilon+\eta-s+\frac{2\epsilon-\eta+s}{1-\epsilon-s}\cr && \approx 2\epsilon^2+3\epsilon s-\epsilon \eta -\eta s +s^2.
\end{eqnarray}
The variation is therefore of second order in the slow-roll parameters, and any corrections to the observables will vanish to lowest order in slow roll. As expected, the above analysis reduces to the simpler case where $c_{s}=1$ and $s=0$ presented in \cite{Kinney:2005vj}. It is therefore clear that even though the sound horizon crossing formalism is exact for the de Sitter case and approximate in the slow roll case, one must be careful when applying it to more exotic models. If the quantum modes do not freeze after sound horizon crossing, this will manifest itself in Eq. (\ref{eq:y formalism 4}) as a strong deviation from zero. Viable inflation models for which the horizon crossing formalism fails, have been constructed for canonical scalar fields \cite{Kinney:2005vj,Tzirakis:2007bf}. In the next section we elaborate on a DBI inflation model which was first introduced by Spalinski \cite{Spalinski:2007un} and is characterized by the same behavior. For the remaining of the paper, primes denote derivatives with respect to the field $\phi$, and overdots derivatives with respect to the coordinate time $t$.

\section{Ultra-slow roll DBI inflation}
\label{sec:Ultra-slow roll DBI inflation}

In this section we discuss the case of a flat potential, $V(\phi)=V_{0}={\mathrm const.}$, for an arbitrary speed of sound. The simpler case of a flat potential for a canonical scalar field was studied in \cite{Kinney:2005vj,Tsamis:2003px}, under the term ``ultra-slow roll'' inflation which can briefly be summarized as follows: The equation of motion for the inflaton field
\begin{equation}
\label{EoM1}
\ddot \phi + 3 H \dot \phi + V^{\prime} (\phi)=0,
\end{equation}
with 
\begin{equation}
\label{eta1}
\eta=-\frac{\ddot \phi}{H \dot \phi},
\end{equation}
can be expressed in terms of the flow parameter $\eta$ as
\begin{equation}
\label{EoM2a}
\eta-3-\frac{V'(\phi)}{H \dot \phi}=0.
\end{equation}
In the case of a flat potential, the equation of motion can be expressed solely in terms of $\eta$ as
\begin{equation}
\label{EoM3}
\eta=3.
\end{equation}
Even though the slow-roll approximation is never valid, since $\eta=3$, and the horizon crossing formalism is strongly violated, it was shown in \cite{Kinney:2005vj} that $\epsilon\rightarrow 0$ in the late-time limit, and the equation of motion for the quantum modes is identical to the equation of motion in de Sitter space, resulting in a scale invariant spectrum.

The more general case of a flat potential with arbitrary but constant speed of sound was first introduced by Spalinski \cite{Spalinski:2007un}. For a flat potential in DBI, the Hamilton-Jacobi equation becomes
\begin{equation}
\label{HJ 1}
H^2(\phi)-\frac{4M_{P}^2}{3(\gamma+1)}\left[H'(\phi)\right]^2=\frac{V_{0}}{3M_{P}^2}.
\end{equation}
Since the potential is constant, the friction term will dominate at late time causing the field to come to a stop at some field value $\phi_{0}$, which can be set to zero without loss of generality. We can then solve exactly for the evolution of the Hubble and the flow parameters on a flat potential and find that
\begin{eqnarray}
\label{H and flow evolution}
H(\phi)&=&\sqrt{\frac{V_{0}}{3M_{P}^2}}\cosh\left(\frac{\sqrt{3(\gamma+1)}}{2}\frac{\phi}{M_{P}}\right),\cr \epsilon(\phi)&=&\frac{3(\gamma+1)}{2\gamma}\tanh^2\left(\frac{\sqrt{3(\gamma+1)}}{2}\frac{\phi}{M_{P}}\right),\cr \xi(\phi)&=&\left(\frac{3(\gamma+1)}{2\gamma}\right)^2\tanh^2\left(\frac{\sqrt{3(\gamma+1)}}{2}\frac{\phi}{M_{P}}\right),\cr \eta(\phi)&=&\frac{3(\gamma+1)}{2\gamma}.
\end{eqnarray}
This is the solution that was presented in \cite{Spalinski:2007un}. The early and late-time values of the flow parameters are summarized in the following table. 

\begin{table}[htbp]
\centering
\caption{Ultra-slow roll DBI inflation}\label{t1}
\begin{tabular}{|c||c|c|} \hline
& early time ($\phi \rightarrow \infty$) & late time ($\phi \rightarrow 0$)   \\ \hline \hline
$\epsilon(\phi)$ & $3(\gamma+1)/(2\gamma)$ & $0$ \\ \hline
$\eta(\phi)$ & $3(\gamma+1)/(2\gamma)$ & $3(\gamma+1)/(2\gamma)$ \\ \hline
$\xi(\phi)$ & $[3(\gamma+1)/(2\gamma)]^2$ & $0$ \\ \hline
\end{tabular}
\end{table}

We can equivalently recover the above evolution by expressing the equation of motion for DBI inflation in terms of the flow parameters, as was done in Eq. (\ref{EoM2a}) for the case of a canonical scalar field. After tedious but straightforward algebra it can be shown that the equation of motion for DBI inflation (\ref{eq: EOM for DBI}) can be written as
\begin{equation}
\label{eq:DBI EoM}
2\left(\frac{\gamma}{\gamma+1}\right)\eta-\left(\frac{\gamma}{\gamma+1}\right)^2s-3-\frac{V'(\phi)}{H\dot \phi \gamma}=0,
\end{equation}
where it should be noted that the above result is exact, since no assumptions of slow roll have been made, and it also reduces to Eq. (\ref{EoM2a}) in the case where $\gamma=1$ and $s=0$. For the evolution considered in this section, we find that the flow parameter $\eta$ which corresponds to a flat potential ($V'(\phi)=0$) with constant speed of sound ($s=0$) according to Eq. (\ref{eq:DBI EoM}) is given by 
\begin{equation}
\label{eq:eta for flat pot}
\eta=\frac{3(\gamma+1)}{2\gamma},
\end{equation}
which as expected, is exactly the result that was found starting from the Hamilton-Jacobi equation.

We next  calculate the power spectrum of curvature perturbations $P_{\mathcal{R}}$ for the late-time evolution. Since the speed of sound is constant, $s$ will vanish and using the values of the flow parameters in the late-time limit from table \ref{t1}, we can write the mode equation (\ref{eq:exactmode}) as
\begin{equation}
\label{eq:exactmode flat V 1}
y^2 \frac{d^2 u_k}{dy^2} + \left[y^2 - \left(2-\frac{9}{4}\left[1-\gamma^{-2}\right]\right)\right] u_k = 0.
\end{equation}
If we now set 
\begin{equation}
\alpha=\frac{9}{4}\left(1-\frac{1}{\gamma^2}\right),
\end{equation}
the mode equation becomes
\begin{equation}
\label{eq:exactmode flat V 2}
y^2 \frac{d^2 u_k}{dy^2} + \left[y^2 - \left(2-\alpha\right)\right] u_k = 0,
\end{equation}
which can easily be solved as
\begin{equation}
\label{eq:sol of exactmode flat V 1}
u_{k} =\frac{1}{2}\sqrt{\frac{\pi}{c_{s}k}} \sqrt{y}H_{\nu}(y),
\end{equation}
where
\begin{equation}
\label{eq:nu of exactmode flat V 1}
\nu=\frac{3}{2}\sqrt{1-\frac{4}{9}\alpha}=\frac{3}{2\gamma}.
\end{equation}

It should be noted that Eq. (\ref{eq:exactmode flat V 2}) which describes the evolution of the quantum modes $u_{k}$ for the case of a flat potential with constant speed of sound, is identical to Eq. (117) of \cite{Tzirakis:2007bf} which describes the evolution of the modes for a canonical scalar field evolving on both branches of a tree-level hybrid potential \cite{Linde:1993cn} in the small $\epsilon$ limit. This is an interesting duality between these two different classes because they both appear to be degenerate in terms of observables, since they both predict negligible tensors in the small $\epsilon$ limit, and non-Gaussian signatures, as we show below.

The power  spectrum of curvature perturbations is
\begin{equation}
\label{P(k) u small epsilon 3}
P_{\cal R}^{1/2}\left(k\right) = \sqrt{\frac{k^3}{2 \pi^2}} \left\vert\frac{u_k}{z}\right\vert_{y\rightarrow 0} = 2^{\nu-3/2}\frac{\Gamma(\nu)}{\Gamma(3/2)} \frac{H^2}{2\pi \dot\phi} y^{3/2-\nu},
\end{equation}
where the power spectrum was evaluated in the long wavelength limit. The scalar spectral index is also given by
\begin{equation}
\label{spec ind 1}
n_{s}-1=\left. \frac{d \ln P_{\mathcal R}}{d\ln k}\right|_{\gamma aH={\mathrm const.}}=3-2\nu,
\end{equation}
or
\begin{equation}
\label{spec ind 2}
n_{s}-1=3\left(1-\frac{1}{\gamma}\right).
\end{equation}
Some remarks should be made at this point. The evolution studied in this section, that of a flat potential with constant speed of sound, predicts a blue spectrum for any value of $\gamma$ which  respects causality and is different than unity, and is therefore disfavored by observation. The tilt of the spectrum varies from extremely blue in the case where $\gamma \rightarrow \infty$, to a scale invariant spectrum when $\gamma=1$. It is also obvious from the above result that a nearly scale invariant spectrum can only happen in the case where the speed of sound is close to unity, and will correspond to very small levels of non-Gaussian signatures. The observational degeneracy therefore between this model and hybrid inflation is evident.

We end this section by examining the validity of the sound horizon crossing formalism. In the limit where $\epsilon$ is small, Eq. (\ref{eq:y formalism 4}) reduces to
\begin{equation}
\label{eq:y formalism 6}
\frac{d}{d{\rm ln}y}\left[{\rm ln}\left(\frac{H}{\sqrt{c_{s}\epsilon}}y^{3/2-\nu}\right)\right]= \frac{3}{2}-\nu-\eta,
\end{equation}  
and using Eqs. (\ref{eq:eta for flat pot}) and (\ref{eq:nu of exactmode flat V 1}) we find that
\begin{equation}
\label{eq:y formalism 7}
\frac{d}{d{\rm ln}y}\left[{\rm ln}\left(\frac{H}{\sqrt{c_{s}\epsilon}}y^{3/2-\nu}\right)\right]= -\frac{3}{\gamma}.
\end{equation}  
From the above result we therefore conclude that any value of $\gamma$ that is not extremely large, will correspond to quantum modes that do not freeze when they exit the sound horizon, and the sound horizon crossing formalism will then fail. This can also be seen by calculating the scalar spectral index using the sound horizon crossing formalism which can be shown to give the following wrong result
\begin{equation}
\label{spec ind 3}
\left. \frac{d \ln P_{\mathcal R}}{d\ln k}\right|_{k=\gamma aH}=3\left(1+\frac{1}{\gamma}\right),
\end{equation}
as does the usual slow roll expression
\begin{equation}
\label{spec ind 4}
\left.n_{s}-1\right \vert_{SR}=-4\epsilon+2\eta-2s =  3\left(1+\frac{1}{\gamma}\right).
\end{equation}

The statement therefore that one is allowed to relate physical quantities like the curvature perturbations far outside the sound horizon, to conserved quantities which are evaluated when they exit the sound horizon is not always true. One should hence be careful when trying to construct inflationary models that are characterized by non-trivial field evolutions like the case of a flat potential presented here, a field rolling up an inverted potential \cite{Tzirakis:2007bf}, or the case where the inflationary potential has features \cite{Adams:2001vc,Starobinsky:1992ts}.

\section{Isokinetic inflation}

In this section we study an interesting class of models characterized by a constant field velocity $\dot\phi = {\mathrm const.}$ Such models have the useful property that it is possible to solve exactly for the background field evolution in both the canonical and DBI cases, although the slow-roll approximation must be used for the perturbations. The case of isokinetic inflation with a canonical Lagrangian is familiar: ``chaotic'' inflation generated by a quadratic potential,
\begin{equation}
V\left(\phi\right) = m^2 \phi^2.
\end{equation}
In the slow-roll limit, since the Hubble parameter $H \propto \sqrt{V}$, we have that 
\begin{equation}
\dot\phi \propto H'\left(\phi\right) \simeq {\mathrm const.}
\end{equation} 
This can be generalized to an exactly solvable model as follows:
For canonical scalar fields, constant field velocity can only be achieved when 
\begin{equation}
\label{eta constant vel}
\eta(\phi)=0,
\end{equation}
at all times as it can be seen from Eq. (\ref{eta1}). From Eqs. (\ref{eq: intro flow 1}) it is then clear that all the higher order flow parameters will also vanish
\begin{equation}
\label{flow for const vel}
^{\ell}\lambda(\phi)=0 \hspace{0.5cm} {\rm for} \hspace{0.5cm} \ell=1,2,...
\end{equation}
The only non-zero flow parameter is therefore $\epsilon$ which can be expressed in terms of the field value as follows: from the Hamilton-Jacobi formalism we have that
\begin{equation}
\label{HJ for eps}
H'(\phi)=-\frac{\dot \phi}{2M_{P}^2} \equiv B = {\mathrm const.},
\end{equation}
which can easily be integrated in order to give that
\begin{equation}
\label{H for const vel 1}
H(\phi)=B\phi +(H_{0}-B\phi_{0}),
\end{equation}
for some initial values $\phi_{0}$ and $H_{0}$. If we next set 
\begin{equation}
H_{0}-B\phi_{0}=C,\nonumber 
\end{equation}
we find that
\begin{equation}
\label{H for const vel 2}
H(\phi)=B\phi +C.
\end{equation}
Using finally Eqs. (\ref{eq: intro flow 1}) we obtain the following exact result for the evolution of $\epsilon$ in terms of the field value
\begin{equation}
\label{eps for const vel 1}
\epsilon(\phi)=\frac{2M_{P}^2}{\left(\phi+C/B\right)^2},
\end{equation}
which is positive-definite as expected. We can also find an exact expression for the potential $V(\phi)$ that can support the above field evolution. From Eq. (\ref{eq: V for canonical}) and using Eqs. (\ref{H for const vel 2}) and (\ref{eps for const vel 1}) we have
\begin{equation}
\label{V for const vel 1}
V(\phi)=3M_{P}^2B^2\left[\left(\phi+\frac{C}{B}\right)^2-\frac{2M_{P}^2}{3}\right].
\end{equation}
For $\phi$ large, this reduces as expected to $V\left(\phi\right) \simeq m^2 \phi^2$, with $m^2 = 3 M_P^2 B^2$, and the corresponding observables are the same. 

The non-canonical generalization of this case is more interesting. We start by identifying the relation between the flow parameters which correspond to such evolution. It can be shown that the generalization of Eq. (\ref{eta1}) in the case where the speed of sound is an arbitrary function of time is given by
\begin{equation}
\label{ddot phi 1}
\ddot \phi = -\dot \phi H(\eta-s),
\end{equation}
and constant field velocity can only take place if
\begin{equation}
\label{eta equal s}
\eta(\phi)=s(\phi).
\end{equation}
The above result can equivalently be expressed in terms of $H(\phi)$ and $\gamma(\phi)$ as
\begin{equation}
\label{H and gamma 1}
\frac{H''(\phi)}{H'(\phi)}=\frac{\gamma'(\phi)}{\gamma(\phi)}.
\end{equation}
Using Eqs. (\ref{eq:flowparams}) we  also find the following exact relation between higher order flow parameters
\begin{equation}
\label{xi equal eps rho}
\xi(\phi)=\epsilon(\phi) \rho(\phi),
\end{equation}
which can again be translated in the following relation between $H(\phi)$ and $\gamma(\phi)$,
\begin{equation}
\label{H and gamma 2}
\frac{H'''(\phi)}{H'(\phi)}=\frac{\gamma''(\phi)}{\gamma(\phi)}.
\end{equation}
Equations (\ref{H and gamma 1}) and (\ref{H and gamma 2}) start revealing a pattern between the $(\ell+1)$-th derivative of $H(\phi)$ and the $\ell$-th derivative of $\gamma(\phi)$. Indeed, using the method of mathematical induction it can be shown that
\begin{equation}
\label{H and gamma 3}
\frac{1}{H'}\left(\frac{d^{\ell+1}H}{d\phi^{\ell+1}}\right)=\frac{1}{\gamma}\left(\frac{d^{\ell}\gamma}{d\phi^{\ell}}\right) \hspace{0.5cm} {\rm for} \hspace{0.5cm} \ell=1,2,...,
\end{equation}
which can also be expressed in terms of the flow parameters as 
\begin{equation}
\label{Htower gammatower}
^{\ell+1}\lambda(\phi)=^{\ell}\alpha(\phi) \epsilon(\phi), \hspace{0.2cm} \ell=1,2,...
\end{equation}
The parameter $\epsilon$ therefore plays the role of a bridge between the $(\ell+1)$-th element of the $H$-tower and the $\ell$-th element of the $\gamma$-tower, and we can then express the elements of both towers in terms of the elements of the $H$-tower only.

We can also express the flow parameters $\epsilon\left(\phi\right)$ and $\gamma\left(\phi\right)$ in terms of $H\left(\phi\right)$ by taking
\begin{equation}
\dot\phi = - \frac{2 M_P^2}{\gamma\left(\phi\right)} H'\left(\phi\right),
\end{equation}
so that
\begin{equation}
\gamma\left(\phi\right) = H'\left(\phi\right) / B,
\end{equation}
where, as in the canonical case,
\begin{equation}
B \equiv - \frac{\dot\phi}{2 M_P^2}.
\end{equation}
The parameter $\epsilon$ is then given by
\begin{equation}
\epsilon\left(\phi\right) = \frac{2 M_P^2}{\gamma\left(\phi\right)} \left(\frac{H'\left(\phi\right)}{H\left(\phi\right)}\right)^2 = 2 M_P^2 B \frac{H'\left(\phi\right)}{H^2\left(\phi\right)}.
\end{equation}
Unlike the  canonical case, if we allow the speed of sound to vary, there is no unique solution for $H\left(\phi\right)$, and therefore isokinetic solutions correspond to a {\it class} of potentials, with
\begin{equation}
\eta(\phi) = s(\phi) = \frac{2 M_P^2 B}{H\left(\phi\right)} \left(\frac{\gamma'\left(\phi\right)}{\gamma\left(\phi\right)}\right).
\end{equation}
The canonical limit corresponds to $\gamma = {\mathrm const.} = 1$, so that $\eta = s = 0$. We proceed by {\it ansatz}: for $\phi$ large, assume that $H\left(\phi\right)$ is dominated by a term of order $\phi^p$, so that 
\begin{equation}
H\left(\phi\right) = B \mu^{\left(1 - p\right)} \phi^p,
\end{equation}
where $\mu$ has dimension of mass. In the slow-roll (large $\phi$) limit, this corresponds to a potential dominated by a term of order $V \propto H^2 \propto \phi^{2 p}$, and is therefore a generalization of chaotic, or ``large-field'' inflation for the case of varying sound speed. It is straightforward to evaluate the  parameters,
\begin{equation}
\gamma(\phi) = p \mu^{\left(1 - p\right)} \phi^{\left(p - 1\right)},
\end{equation}
and
\begin{equation}
\epsilon(\phi) = \frac{2 p M_P^2}{\mu^{\left(1 - p\right)}} \phi^{-\left(p + 1\right)} = \left(\frac{\phi_e}{\phi}\right)^{p + 1},
\end{equation}
where inflation ends with $\epsilon\left(\phi_e\right) = 1$ at a field value
\begin{equation}
\label{eq:phiendnc}
\phi_e = \left(\frac{2 p M_P^2}{\mu^{\left(1 - p\right)}}\right)^{1 / \left(p + 1\right)}.
\end{equation}
We can then write $\gamma$ as
\begin{equation}
\label{eq: gamma phi}
\gamma\left(\phi\right) = 2 p^2 \left(\frac{M_P}{\phi_e}\right)^2 \left(\frac{\phi}{\phi_e}\right)^{p - 1}.
\end{equation}
We recover the canonical case $\gamma = 1$, by setting $p = 1$. The number of e-folds $N\left(\phi\right)$ is given by:
\begin{eqnarray}
N\left(\phi\right) &=& \frac{1}{\sqrt{2 M_P^2}} \int_{\phi_e}^{\phi}{\sqrt{\frac{\gamma\left(\phi\right)}{\epsilon\left(\phi\right)}} d \phi}= \frac{p}{\phi_e} \int_{\phi_e}^{\phi}{\left(\frac{\phi}{\phi_e}\right)^p d \phi}\cr
&=& \frac{p}{p + 1} \left(\frac{1}{\epsilon} - 1\right),
\end{eqnarray}
and we can then write $\epsilon$ in terms of the number of e-folds $N$ as
\begin{equation}
\label{eq: eps phi}
\epsilon\left(N\right) = \frac{1}{1 + \left(p + 1\right) N / p}.
\end{equation}
Note that the spectral index $n_{s}$ depends only on $\epsilon$, since in the slow-roll limit,
\begin{equation}
\label{spec ind 1 for slow roll}
n_{s}-1=-4\epsilon+2\eta -2s = -4 \epsilon=- \frac{4}{1+\left(p + 1\right) N/p}.
\end{equation}
Using this expression, for $N = \left[46,60\right]$, and $p = 1, \ldots, \infty$, the spectral index is confined to a narrow range for the entire class of potentials,
\begin{equation}
n_{s} = \left[0.915,0.967\right],
\end{equation}
as  can be seen from Table II. (The case where $\ddot \phi \approx 0$ was considered in \cite{Li:2008qc}, where a red-tilted spectrum was obtained not through the parameter $\epsilon \sim \mathcal{O}(10^{-4})$, but rather through the parameter $\eta$).

\begin{table}[htbp]
\centering
\caption{Values for $n_{s}$ in terms of $p$ and $N$}\label{t2}
\begin{tabular*}{0.35\textwidth}{@{\extracolsep{\fill}} |c||c|c|c|c|} \hline
& $p=1$ & $p=2$ & $p=5$ & $p=\infty$   \\ \hline \hline
$N=46$ & $0.957$ & $0.943$ & $0.929$ & $0.915$ \\ \hline
$N=60$ & $0.967$ & $0.956$ & $0.945$ & $0.934$ \\ \hline
\end{tabular*}
\end{table}

The tensor/scalar ratio $r$ is less constrained,
\begin{eqnarray}
\label{eq: r phi}
r =\frac{16\epsilon}{\gamma} &=& \frac{8}{p^2} \left(\frac{\phi_e}{M_P}\right)^2 \epsilon^{2 p / \left(p + 1\right)}\cr
&=& \frac{8}{p^2} \left(\frac{\phi_e}{M_P}\right)^2 \left(1+\frac{p+1}{p}N\right)^{-2 p / \left(p + 1\right)}
\end{eqnarray}
Here $\phi_e$ is determined by the arbitrary mass scale $\mu$ (\ref{eq:phiendnc}), so $r$ is not determined by the number of e-folds alone as in the canonical case. We can, however, place an upper limit on $r$ by demanding that the theory be causal, that is $c_s = \gamma^{-1} \leq 1$ throughout the inflationary evolution. Since $\gamma$ is decreasing with time, this corresponds to a lower limit on $\gamma$ at the end of inflation
\begin{equation}
\gamma\left(\phi_e\right) = 2 p^2 \left(\frac{M_P}{\phi_e}\right)^2 \geq 1,
\end{equation}
or
\begin{equation}
\label{eq: phi end}
\left(\frac{\phi_e}{M_P}\right) \leq p \sqrt{2}.
\end{equation}
We then have an {\it upper limit} on the tensor/scalar ratio,
\begin{equation}
\label{eq:upper limit for r}
r \leq 16 \epsilon^{2 p / \left(p + 1\right)} = 16  \left(1+\frac{p+1}{p}N\right)^{-2 p / \left(p + 1\right)}.
\end{equation}
The right-hand side is maximized for the canonical case $p = 1$ and $N = 46$, for which $r = 0.17$. In the limit $p \gg 1$ and $N = 46$, the upper limit is much stronger, $r < 0.007$. This is therefore an example of a class of models which predict a spectral index within the observationally favored range, but which can have an arbitrarily low tensor/scalar ratio. The field excursion $\Delta \phi / M_P$ during inflation can be calculated using
\begin{equation}
\frac{\phi\left(N\right)}{\phi_e} = \left(1+\frac{p+1}{p}N\right)^{1 / \left(p + 1\right)},
\end{equation}
so that
\begin{eqnarray}
\frac{\Delta\phi}{M_P} &=& \frac{\phi\left(N\right) - \phi_e}{M_{P}}\cr
&=& \left(\frac{\phi_e}{M_P}\right) \left[\left(1+\frac{p + 1}{p}N\right)^{1 / \left(p + 1\right)} - 1\right]\cr
&=& \frac{p}{4} \sqrt{2r} \left[\left(1+\frac{p + 1}{p}N\right)^{1 / \left(p + 1\right)} - 1\right]\cr &&\times\left(1+\frac{p+1}{p}N\right)^{p/\left(p + 1\right)}.
\end{eqnarray}
We see that for $r \ll 1$, $\Delta\phi \ll M_P$. In this limit, it is possible to embed such potentials into a stringy realization of inflation such as a DBI scenario, since the field variation will be smaller than the typical size of a compactified dimension. 

\begin{figure}[!h]
\includegraphics[angle=0, width=\columnwidth] {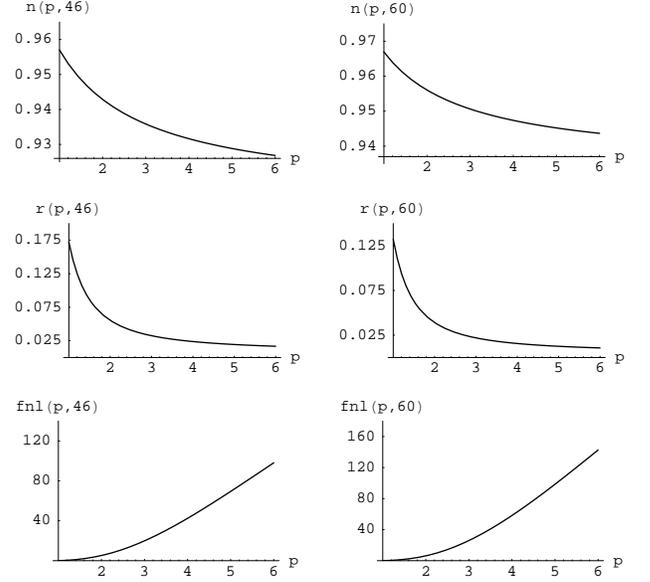}
\caption{The spectral index $n_{s}$ (top two), the upper limit for $r$ (middle two), and the lower limit for $f_{NL}$ (bottom two) for $N=46$ and $N=60$ as a function of $p$.}
\label{fig:Observs}
\end{figure}

The primordial non-Gaussianity can also be expressed in terms of the shape of the potential and the amount of inflation. Using Eqs. (\ref{eq: gamma phi}) and (\ref{eq: eps phi}) we find that
\begin{eqnarray}
\label{eq: fnl 2}
&&f_{NL}^{\rm equil.} = \frac{35}{108}\left[4p^4\left(\frac{M_{P}}{\phi_{e}}\right)^4\left(\frac{\phi}{\phi_{e}}\right)^{2(p-1)}-1\right]\cr &=& \frac{35}{27}\left[p^4\left(\frac{M_{P}}{\phi_{e}}\right)^4\left(1+\frac{p+1}{p}N\right)^{\frac{2(p-1)}{p+1}}-\frac{1}{4}\right]
\end{eqnarray}
exactly. Demanding again that the theory be causal, we can substitute the condition (\ref{eq: phi end}) into the above result, and recover the following {\it lower limit} for the level of non-Gaussianities
\begin{equation}
\label{eq: fnl 3}
f_{NL}^{\rm equil.} \geq  \frac{35}{108}\left[\left(1+\frac{p+1}{p}N\right)^{\frac{2(p-1)}{p+1}}-1\right],
\end{equation}
which equivalently can be expressed as
\begin{equation}
\label{eq: fnl eps}
f_{NL}^{\rm equil.} \geq  \frac{35}{108}\left[\epsilon^{2(1-p)/(p+1)}-1\right].
\end{equation}
The right-hand side is minimized for the canonical case where $p=1$, and is independent of the number of e-folds $N$ since 
\begin{equation}
\label{eq: fnl 4}
f_{NL}^{\rm equil.}(p=1, N) \geq  0,
\end{equation}
for any $N$ value. 
We summarize the previous results in Fig. \ref{fig:Observs}, where we see the general way in which the upper limit for $r$ and the lower limit for $f_{NL}$ depend on $p$. We can then generalize the statement that we mentioned before for the class of models with $p \gg 1$. In addition to the fact that they predict a spectral index within the observationally favored range and arbitrarily low $r$, they also can generate significant levels of non-Gaussian signatures. 

\begin{figure}[!h]
\includegraphics[angle=0, width=\columnwidth] {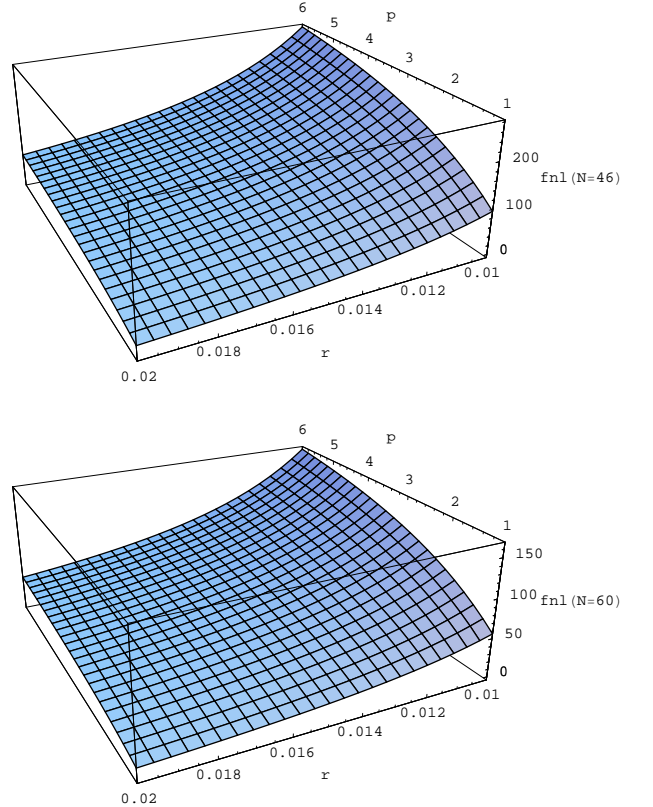}
\caption{$f_{NL}$ as a function of $r$ for different values of $p$.}
\label{fig:specrfnl}
\end{figure}

We next find an expression for $f_{NL}$ in terms of the tensor/scalar ratio $r$. Using Eq. (\ref{eq: r phi}) we have that
\begin{equation}
\label{eq: phiend over Mp}
\left(\frac{M_{P}}{\phi_e}\right)^4=\frac{64}{r^2p^4}\left(1+\frac{p+1}{p}N\right)^{-4 p / \left(p + 1\right)},
\end{equation}
and Eq. (\ref{eq: fnl 2}) can then be written as
\begin{equation}
\label{eq: fnl in terms of r}
f_{NL}^{\rm equil.} =  \frac{35}{108}\left[\frac{256}{r^2}\left(1+\frac{p+1}{p}N\right)^{-2}-1\right].
\end{equation}
It should be noted that this is an exact result and demonstrates that small values of $r$ generate large values of $f_{NL}$. This is also illustrated in Fig. \ref{fig:specrfnl}, where it can be seen that a non-detectable $r$ ($r \leq 0.01$) can potentially correspond to a detectable $f_{NL}$.

\begin{figure*}
\includegraphics[angle=0, width=6.8in] {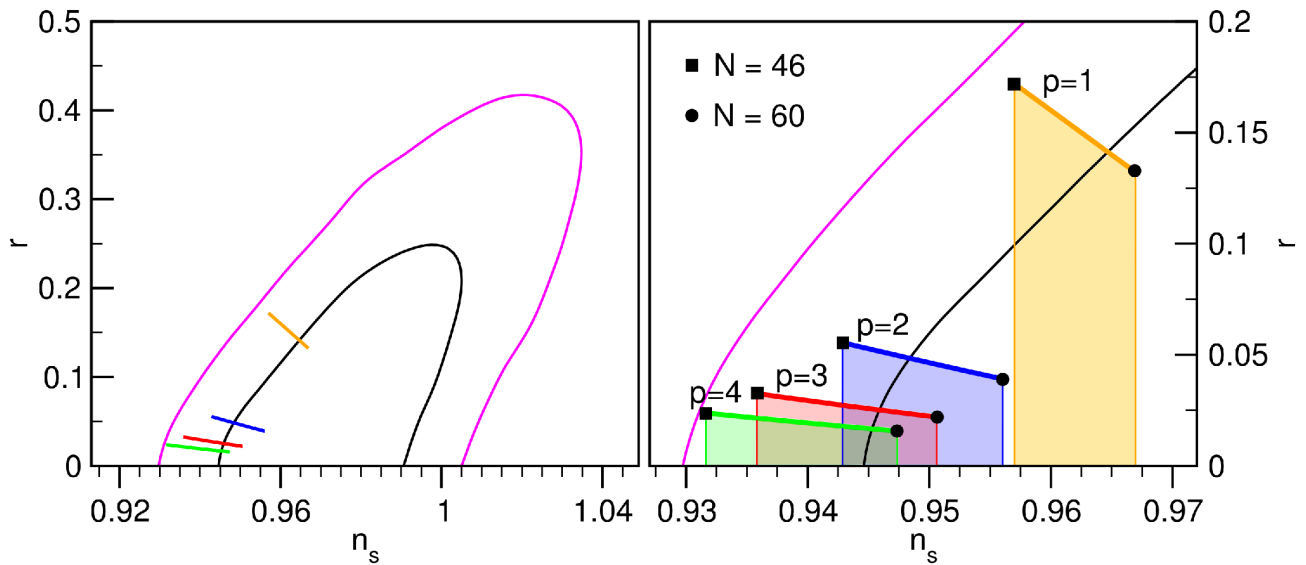}
\caption{68 \% (black) and 95 \% C.L. (magenta) on the $n_{s}$ and $r$ parameter space for WMAP5 alone taken from \cite{Kinney:2008wy}. In the left figure we plotted the predictions of isokinetic inflation for the cases where $p=1$ (orange), $p=2$ (blue), $p=3$ (red), and $p=4$ (green) using the upper limit for $r$ given by Eq. (\ref{eq:upper limit for r}). The right figure is the left one zoomed in to the region of observational interest for these models. The colored shaded regions constitute the observational range for an arbitrary value of $r$. The full squares  (full circles) correspond to the case where a mode crossed the sound horizon $46$  ($60$) e-folds before the end of inflation.}
\label{fig:nrSHADEDfinal}
\end{figure*}

Finally, in Fig. \ref{fig:nrSHADEDfinal} we present the results in the $(n_{s},r)$ plane in terms of the number of e-folds $N$, and for different values of $p$. It can be seen that they all lie within the WMAP5 observationally favored region.

\section{Conclusions}

In this paper, we consider non-canonical generalizations of two interesting classes of canonical inflation models. First, we study ``ultra-slow roll'' inflation, which is a class of inflation models for which the slow-roll approximation is strongly violated, so that quantum modes evolve rapidly on superhorizon scales \cite{Kinney:2005vj,Tzirakis:2007bf}. This scenario generalizes naturally to non-canonical Lagrangians, and exact solutions for the background evolution were found by Spalinski \cite{Spalinski:2007un}. We calculate solutions for the primordial power spectrum in this background, and show that, as with the model's canonical counterpart, the ``horizon crossing'' formalism strongly breaks down. Second, we consider the non-canonical generalization of the simplest ``chaotic'' inflation scenario, with a potential $V\left(\phi\right) = m^2 \phi^2$, for which the field evolves with approximately constant velocity $\dot\phi \simeq {\mathrm const.}$ We find a class of related non-canonical solutions with polynomial potentials $V\left(\phi\right) \propto \phi^p$ and constant field velocity $\dot\phi = {\mathrm const.}$, but with varying speed of sound. Unlike the canonical case, the non-canonical model can have an arbitrarily small tensor/scalar ratio combined with a slightly red-tilted power spectrum, $1 - n \sim 0.05$, consistent with current data. Of particular interest is that this class of models is marked by a correlation between the tensor/scalar ratio and the amplitude $f_{NL}$ of non-Gaussianity, such that parameter regimes with small tensor/scalar ratio have {\it large} associated non-Gaussianity. Such behavior is {\it not} generic: slow-roll inflation models with unobservably low tensor amplitudes also have small non-Gaussianity, so the presence of such a signal is an indication of a varying sound speed during inflation, and therefore presents a useful observational target. 

In this paper, we have not addressed the issue of whether or not potentials of these types could be naturally incorporated into string-based models. Because manifolds in string theory are typically compactified on scales of order $M_P$, inflation models with $\Delta \phi > M_P$ present difficulties for string-based model building, although exceptions based on monodromy have recently been proposed by Silverstein and Westphal \cite{Silverstein:2008sg}. All of the scenarios we consider here have a natural $r \ll 1$ limit, so that $\Delta\phi \ll M_P$, and these issues can be avoided altogether. These results may therefore prove useful for stringy model-building. 

\begin{center}
{\bf ACKNOWLEDGMENTS}
\end{center}
K.T. thanks Brian Powell for many useful discussions. This research is supported in part by the National Science Foundation under grant NSF-PHY-0456777.


\begin{thebibliography}{99}

\bibitem{Guth:1980zm}
   A.~H.~Guth,
   Phys.\ Rev.\ D {\bf 23}, 347 (1981).

\bibitem{Linde:1981mu}
   A.~D.~Linde,
   Phys.\ Lett.\ B {\bf 108}, 389 (1982).

\bibitem{Albrecht:1982wi}
   A.~Albrecht and P.~J.~Steinhardt,
   Phys.\ Rev.\ Lett.\  {\bf 48}, 1220 (1982).


\bibitem{Mukhanov:1981xt}
  V.~F.~Mukhanov and G.~V.~Chibisov,
  JETP Lett.\  {\bf 33} (1981) 532
  [Pisma Zh.\ Eksp.\ Teor.\ Fiz.\  {\bf 33} (1981) 549].

\bibitem{Hawking:1982my}
  S.~W.~Hawking and I.~G.~Moss,
  Nucl.\ Phys.\  B {\bf 224}, 180 (1983).

\bibitem{Starobinsky:1982ee}
  A.~A.~Starobinsky,
  Phys.\ Lett.\  B {\bf 117} (1982) 175.

 \bibitem{Guth:1982ec}
  A.~H.~Guth and S.~Y.~Pi,
  Phys.\ Rev.\ Lett.\  {\bf 49}, 1110 (1982).

 \bibitem{Bardeen:1983qw}
  J.~M.~Bardeen, P.~J.~Steinhardt and M.~S.~Turner,
  Phys.\ Rev.\ D {\bf 28}, 679 (1983).

\bibitem{Starobinsky:1979ty}
  A.~A.~Starobinsky,
  JETP Lett.\  {\bf 30} (1979) 682
  [Pisma Zh.\ Eksp.\ Teor.\ Fiz.\  {\bf 30} (1979) 719].

\bibitem{Starobinsky:1980te}
  A.~A.~Starobinsky,
  Phys.\ Lett.\  B {\bf 91} (1980) 99.

\bibitem{Kachru:2003sx}
  S.~Kachru, R.~Kallosh, A.~Linde, J.~M.~Maldacena, L.~McAllister and S.~P.~Trivedi,
  JCAP {\bf 0310} (2003) 013
  [arXiv:hep-th/0308055].

\bibitem{BlancoPillado:2004ns}
  J.~J.~Blanco-Pillado {\it et al.},
  JHEP {\bf 0411}, 063 (2004)
  [arXiv:hep-th/0406230].

\bibitem{Bond:2006nc}
  J.~R.~Bond, L.~Kofman, S.~Prokushkin and P.~M.~Vaudrevange,
  Phys.\ Rev.\  D {\bf 75}, 123511 (2007)
  [arXiv:hep-th/0612197].

\bibitem{Silverstein:2003hf}
  E.~Silverstein and D.~Tong,
  Phys.\ Rev.\  D {\bf 70}, 103505 (2004)
  [arXiv:hep-th/0310221].


\bibitem{Dodelson:1997hr}
  S.~Dodelson, W.~H.~Kinney and E.~W.~Kolb,
  Phys.\ Rev.\  D {\bf 56}, 3207 (1997)
  [arXiv:astro-ph/9702166].

\bibitem{Kinney:1998md}
  W.~H.~Kinney,
  Phys.\ Rev.\  D {\bf 58}, 123506 (1998)
  [arXiv:astro-ph/9806259].


\bibitem{Spergel:2006hy}
  D.~N.~Spergel {\it et al.}  [WMAP Collaboration],
  Implications for cosmology,''
  Astrophys.\ J.\ Suppl.\  {\bf 170}, 377 (2007)
  [arXiv:astro-ph/0603449].

\bibitem{Alabidi:2006qa}
  L.~Alabidi and D.~H.~Lyth,
  JCAP {\bf 0608}, 013 (2006)
  [arXiv:astro-ph/0603539].

\bibitem{Seljak:2006bg}
  U.~Seljak, A.~Slosar and P.~McDonald,
  galaxy clustering and SN constraints,''
  JCAP {\bf 0610}, 014 (2006)
  [arXiv:astro-ph/0604335].

\bibitem{Kinney:2006qm}
  W.~H.~Kinney, E.~W.~Kolb, A.~Melchiorri and A.~Riotto, 
  Phys.\ Rev.\  D {\bf 74}, 023502 (2006)
  [arXiv:astro-ph/0605338].

\bibitem{Martin:2006rs}
  J.~Martin and C.~Ringeval,
  JCAP {\bf 0608}, 009 (2006)
  [arXiv:astro-ph/0605367].

\bibitem{Komatsu:2008hk}
  E.~Komatsu {\it et al.}  [WMAP Collaboration],
  arXiv:0803.0547 [astro-ph].

\bibitem{Dunkley:2008ie}
  J.~Dunkley {\it et al.}  [WMAP Collaboration],
  arXiv:0803.0586 [astro-ph].

\bibitem{Kinney:2008wy}
  W.~H.~Kinney, E.~W.~Kolb, A.~Melchiorri and A.~Riotto,
  arXiv:0805.2966 [astro-ph].

\bibitem{Lorenz:2008je}
  L.~Lorenz, J.~Martin and C.~Ringeval,
  arXiv:0807.2414 [astro-ph].

\bibitem{Lorenz:2008et}
  L.~Lorenz, J.~Martin and C.~Ringeval,
  Phys.\ Rev.\  D {\bf 78}, 083513 (2008)
  [arXiv:0807.3037 [astro-ph]].

\bibitem{Alishahiha:2004eh}
  M.~Alishahiha, E.~Silverstein and D.~Tong,
  Phys.\ Rev.\  D {\bf 70}, 123505 (2004)
  [arXiv:hep-th/0404084].

\bibitem{Chen:2006nt}
  X.~Chen, M.~x.~Huang, S.~Kachru and G.~Shiu,
  JCAP {\bf 0701}, 002 (2007)
  [arXiv:hep-th/0605045].

\bibitem{Spalinski:2007qy}
  M.~Spalinski,
  Phys.\ Lett.\  B {\bf 650}, 313 (2007)
  [arXiv:hep-th/0703248].

\bibitem{Bean:2007eh}
  R.~Bean, X.~Chen, H.~V.~Peiris and J.~Xu,
  arXiv:0710.1812 [hep-th].

\bibitem{LoVerde:2007ri}
  M.~LoVerde, A.~Miller, S.~Shandera and L.~Verde,
  arXiv:0711.4126 [astro-ph].

\bibitem{Lidsey:1995np}
  J.~E.~Lidsey, A.~R.~Liddle, E.~W.~Kolb, E.~J.~Copeland, T.~Barreiro and M.~Abney,
  Rev.\ Mod.\ Phys.\  {\bf 69}, 373 (1997)
  [arXiv:astro-ph/9508078].

\bibitem{Liddle:1994dx}
  A.~R.~Liddle, P.~Parsons and J.~D.~Barrow,
  Phys.\ Rev.\  D {\bf 50}, 7222 (1994)
  [arXiv:astro-ph/9408015].

\bibitem{Kinney:2002qn}
  W.~H.~Kinney,
  Phys.\ Rev.\  D {\bf 66}, 083508 (2002)
  [arXiv:astro-ph/0206032].

\bibitem{Schwarz:2001vv}
  D.~J.~Schwarz, C.~A.~Terrero-Escalante and A.~A.~Garcia,
  Phys.\ Lett.\  B {\bf 517}, 243 (2001)
  [arXiv:astro-ph/0106020].

\bibitem{Peiris:2007gz}
  H.~V.~Peiris, D.~Baumann, B.~Friedman and A.~Cooray,
  Phys.\ Rev.\  D {\bf 76}, 103517 (2007)
  [arXiv:0706.1240 [astro-ph]].

\bibitem{Bean:2008ga}
  R.~Bean, D.~J.~H.~Chung and G.~Geshnizjani,
  arXiv:0801.0742 [astro-ph].

\bibitem{Agarwal:2008ah}
  N.~Agarwal and R.~Bean,
  arXiv:0809.2798 [astro-ph].

\bibitem{Kinney:2005vj}
  W.~H.~Kinney,
  Phys.\ Rev.\  D {\bf 72}, 023515 (2005)
  [arXiv:gr-qc/0503017].

\bibitem{Tzirakis:2007bf}
  K.~Tzirakis and W.~H.~Kinney,
  Phys.\ Rev.\  D {\bf 75}, 123510 (2007)
  [arXiv:astro-ph/0701432].

\bibitem{Spalinski:2007un}
  M.~Spalinski,
  JCAP {\bf 0804}, 002 (2008)
  [arXiv:0711.4326 [astro-ph]].

\bibitem{Muslimov:1990be}
  A.~G.~Muslimov,
  Class.\ Quant.\ Grav.\  {\bf 7}, 231 (1990).

\bibitem{Salopek:1990jq}
  D.~S.~Salopek and J.~R.~Bond,
  Phys.\ Rev.\ D {\bf 42}, 3936 (1990).

\bibitem{Kinney:2005in}
  W.~H.~Kinney and A.~Riotto,
  JCAP {\bf 0603}, 011 (2006)
  [arXiv:astro-ph/0511127].

\bibitem{Maldacena:2002vr}
  J.~M.~Maldacena,
  JHEP {\bf 0305}, 013 (2003)
  [arXiv:astro-ph/0210603].

\bibitem{Acquaviva:2002ud}
  V.~Acquaviva, N.~Bartolo, S.~Matarrese and A.~Riotto,
  Nucl.\ Phys.\  B {\bf 667}, 119 (2003)
  [arXiv:astro-ph/0209156].

\bibitem{Klebanov:2000hb}
  I.~R.~Klebanov and M.~J.~Strassler,
  JHEP {\bf 0008}, 052 (2000)
  [arXiv:hep-th/0007191].


\bibitem{Lidsey:2007gq}
  J.~E.~Lidsey and I.~Huston,
  JCAP {\bf 0707}, 002 (2007)
  [arXiv:0705.0240 [hep-th]].

\bibitem{Chen:2004gc}
  X.~Chen,
  Phys.\ Rev.\  D {\bf 71}, 063506 (2005)
  [arXiv:hep-th/0408084].

\bibitem{Chen:2005ad}
  X.~Chen,
  JHEP {\bf 0508}, 045 (2005)
  [arXiv:hep-th/0501184].

\bibitem{Spalinski:2007kt}
  M.~Spalinski,
  JCAP {\bf 0704}, 018 (2007)
  [arXiv:hep-th/0702118].

\bibitem{Garriga:1999vw}
  J.~Garriga and V.~F.~Mukhanov,
  Phys.\ Lett.\  B {\bf 458}, 219 (1999)
  [arXiv:hep-th/9904176].

\bibitem{Kinney:2007ag}
  W.~H.~Kinney and K.~Tzirakis,
  Phys.\ Rev.\  D {\bf 77}, 103517 (2008)
  [arXiv:0712.2043 [astro-ph]].

\bibitem{Tsamis:2003px}
  N.~C.~Tsamis and R.~P.~Woodard,
  Phys.\ Rev.\  D {\bf 69}, 084005 (2004)
  [arXiv:astro-ph/0307463].

\bibitem{Linde:1993cn}
  A.~D.~Linde,
  Phys.\ Rev.\ D {\bf 49}, 748 (1994)
  [arXiv:astro-ph/9307002].

\bibitem{Adams:2001vc}
  J.~A.~Adams, B.~Cresswell and R.~Easther,
  Phys.\ Rev.\  D {\bf 64}, 123514 (2001)
  [arXiv:astro-ph/0102236].

\bibitem{Starobinsky:1992ts}
  A.~A.~Starobinsky,
  JETP Lett.\  {\bf 55} (1992) 489
  [Pisma Zh.\ Eksp.\ Teor.\ Fiz.\  {\bf 55} (1992) 477].

\bibitem{Li:2008qc}
  M.~Li, T.~Wang and Y.~Wang,
  JCAP {\bf 0803}, 028 (2008)
  [arXiv:0801.0040 [astro-ph]].

\bibitem{Silverstein:2008sg}
  E.~Silverstein and A.~Westphal,
  arXiv:0803.3085 [hep-th].



\end{thebibliography}
\end{document}